# *Effective City Planning: A Data Driven Analysis of Infrastructure and Citizen Feedback in Bangalore*

*Research Article*

*Srishti Mishra**
*PES University*

*Srinjoy Das**
*PES University*

**Both authors contributed equally to this work*

***Abstract*—Analyzing civic data can present a clear picture of the priorities, current performance, and the weaknesses of the city. Civic data can be divided into three main categories - spending, infrastructure and citizen feedback. Analyzing this data can capture the current issues faced by citizens, disparity between government spending and quality of work, and can help in providing effective solutions to these problems. City infrastructure in terms of footpaths, lighting, and parks, as discussed below, provide an insight to the living quality of citizens and can be compared to the annual spending in these sectors. Analysis of complaints and citizen feedback ensure that the sentiments of the citizens are taken into account during planning and can help in pinpointing the areas which need to be improved in the coming future. Continuous analysis and review will help in improving performance of municipal corporations and ensure transparency between citizens and the city officials, resulting in effective city planning.**

**Ranking constituencies across the city based on an analysis of infrastructure across parks, footpaths, and street lighting highlighted that all constituencies rank low in certain areas. The results indicated a low importance towards greenery in terms of Parks, where each constituency has less than 2% of their area as a park. The current population in these constituencies are high and are growing, hence this is a pressing problem and will worsen in the coming years. Comparing the results with the complaints across the city, we found some interesting insights and patterns. Analysis and rankings of footpaths in constituencies were contrary to the number of complaints in these constituencies, with high ranking constituencies (better scores) receiving the highest number of complaints. This could be due to a number of factors and requires further analysis. In terms of street lights, the areas which had low quality street lights were associated with a large number of complaints from citizens, indicating that action needs to be taken immediately. Overall analysis of complaints across constituencies reflected the everyday struggles of the city with the top words being about roads and vehicles. A word cloud analysis demonstrated that the keywords "footpaths" and "garbage" followed "roads", which are both critical problems in Bangalore City today. The source code of the data analysis is linked here https://github.com/srishti-56/city-analytics.**

## I. Introduction

In the past several years, data storage and collection has turned electronic, and there is a vast sea of civic data slowly building up. This data includes city contracts, tenders, bills, pending work, maintenance of city infrastructure, and most importantly citizen complaints. By analyzing the data [8], we can make useful inferences and identify key areas to focus on in the coming future.

In the recent past, using analytics to drive city development, infrastructure and budgets has become more popular and has seen notable successes, namely in Baltimore, Chicago and Philadelphia in the USA. Analytics played a big role through important projects such as deployment of police forces in the city (Philadelphia) [3], abating the rodent issue (Chicago) [2], measuring cleanliness across the city (Los Angeles) [4] etc. Following their example, other cities in the USA are setting up their own analysis-driven solutions. However, every city is unique; in the problems it faces, in its citizen demographics and its major industries. Hence, a specialized and focussed approach is required to understand, analyze and set up a progressive solution with city officials and data analysts. Performance measures and metrics play an important role in assessing and communicating the current state of the city to internal and external stakeholders, that is, the government and its citizens, in order to take effective action. For instance, creating indexes across the city for water availability, healthcare, green space/parks, sanitary facilities, road infrastructure and so on can help measure the effectiveness of municipal work and point out areas for improvement. In addition, analysis of citizen inputs and complaints based on location, priority level, and frequency can help measure citizen priorities versus municipal focus in the area. The Swachh City platform [6] in India collects cleanliness complaints across the city and works with the municipal offices to rectify the problems. A metric measures user happiness and user engagement from data collected through their app. Using location-based complaints from citizens can help measure citizen priorities versus municipal focus in the area, and can measure citizen satisfaction across wards in the city. Integrating an analytical approach with real-time data can ensure speedy progress in terms of repairing city infrastructure, measuring cleanliness, power-saving and so on based on real-time citizen issues.

## II. Data and Approach

In order to harness the power of analytics in City Planning, accurate civic data needs to be collected. The civic data dataset [4] contains general information such as area, population, increase in population etc, as well as details on civic infrastructure for all wards across the city. The complaints data consists of a log of complaints from each ward along with categories, subcategories and descriptions of the issue.

The comprehensive 5-step plan described below is to assess and form key inferences from our civic data.

1. Identifying the current condition of the city, in terms of services and infrastructure across constituencies is the first step in city planning.
2. The next step is to determine the major areas of citizen dissatisfaction in order to correlate these to current conditions and to devise plans to improve the city.
3. Analyze the current expenditure and work done by contractors along with the constituency rankings and citizen complaints.
4. Combine analysis from the constituency infrastructure and service rankings, citizen complaints and current expenditure and contractors to measure quality of work, areas of improvement and where there has been overspending and underspending.
5. Examine the results of the analysis, receive feedback from city officials/civic organizations and improve the analysis. Also investigate the effects of any external factors which have not been considered at the moment.

Bangalore city is divided into around 200 unique wards. The large number of wards made ranking and visualizing the data messy and unreadable. Coalescing these wards into their respective constituencies brought down the count of divisions of the city to around 30 constituencies, as shown in Figure 1, and provided better results.

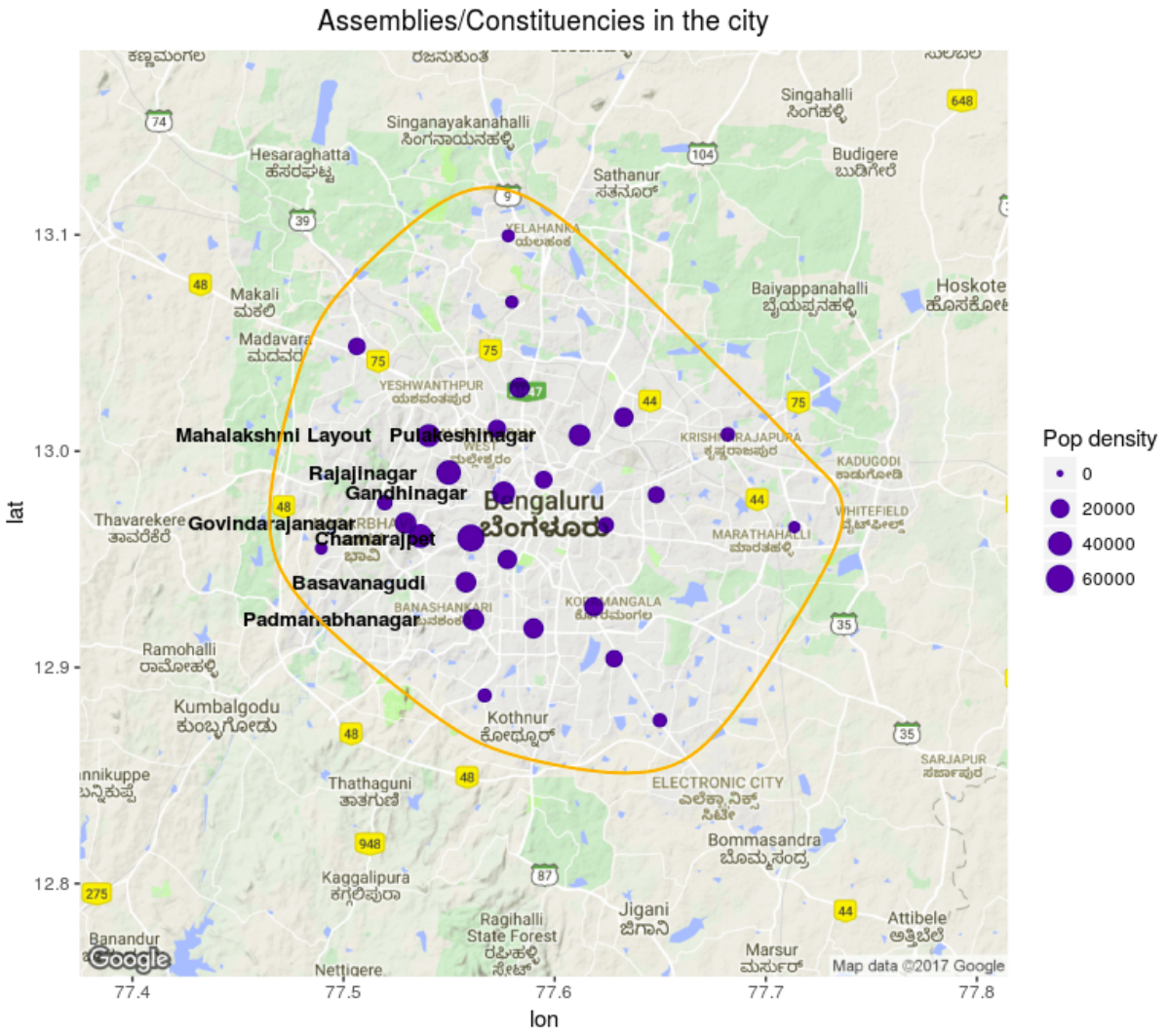

Figure 1. Assemblies/ Constituencies in Bangalore City

Note that the terms “wards”, “assemblies” and “constituencies” are used interchangeably from this point on, and refer to the 30 unique constituencies the city is divided into. The subsequent sections explore and analyze the city infrastructure and citizen complaints.

## III. Analysis of City Infrastructure

The data analyzed on civic infrastructure covers details of Parks, Street Lighting, Footpaths, Bus Stops and Complaints.

### A. *Quality of Parks*

The dataset describes the area of the park, as well as the number of lights and the water source. However, the lighting and water features were dropped as over 60% of the data was missing. Since the quality of each park is being measured, filling in the missing values through normal techniques like averaging, interpolation etc, would severely distort results. This is not acceptable in terms of city planning since these results may be used to help allocate city budgets.

Parks are both a part of citizen services and infrastructure, and the quality of parks across the city is measured with respect to the number of people it’s shared by and the percentage of area it occupies out of the total constituency area.

The results were surprising, less than 2% of the area of the constituency is occupied by parks as seen in Figure 2. With the increase in population, based on census data, and urbanization, the problem will get worse in the coming years. The lack of open air and the reduced quality of air may cause health problems in citizens later on, and hence this should be a priority for the government. Additionally, it was observed that there is barely a square km of park area available for each hundred thousand people living in these constituencies.

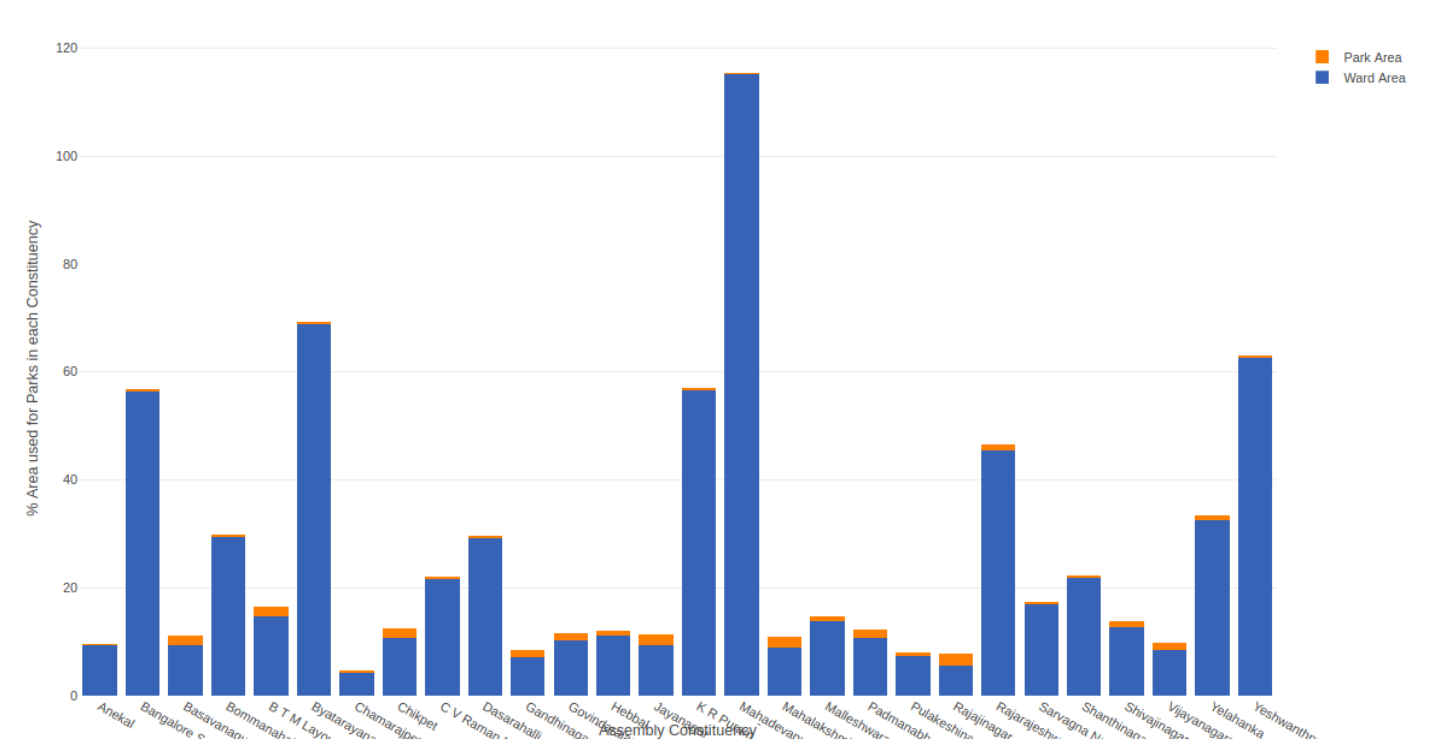

Figure 2. % of Park Area (orange) vs Ward Area (blue) across the city

### B. *Quality of Footpaths*

Constituencies were ranked across the city based on footpath quality; computed as the percentage of the length of usable footpath by total length of the footpath. Comparing these results with the number of complaints revealed a substantial gap between the quality of the footpaths calculated and the citizen feedback, as shown in Figure 3. The analysis indicated that the highest numbers of complaints fell in the highest ranking constituencies. This could mean that either the measurements of usable footpaths are no longer accurate, or that there are temporary factors that are affecting the quality of the footpaths, including issues such as dumping garbage or digging. Further investigation is required to ascertain the cause of this discrepancy.

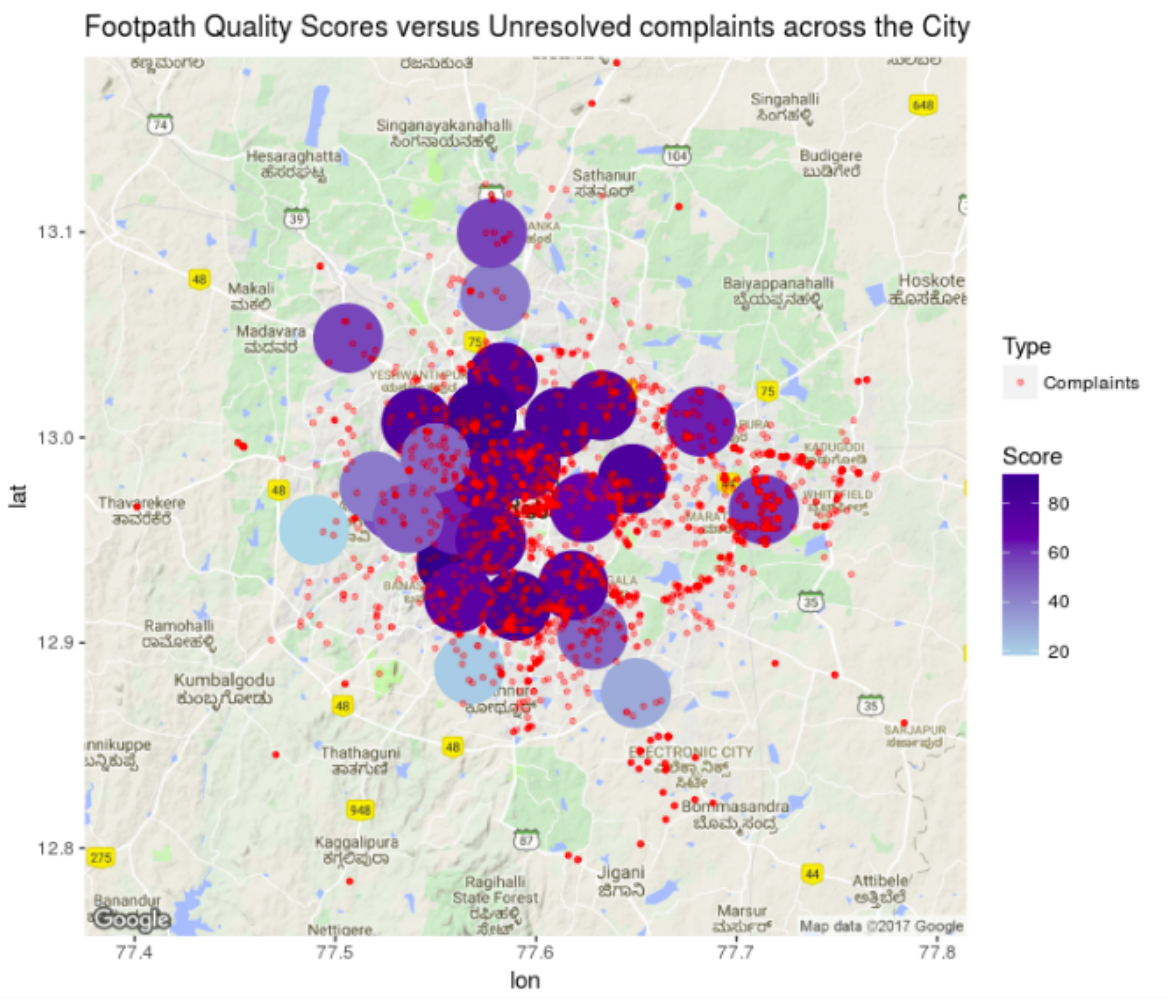

Figure 3. Footpath quality scores Vs. Unresolved complaints across the city.

### C. *Street lighting Quality*

The quality of the street lights in a constituency is measured as the average lux available as compared to the minimum lux safety benchmark. Analyzing the quality of street lights across constituencies with complaints, as shown in Figure 4, exhibited a strong correlation between the complaints and the constituencies' scores. Areas with lower scores had more complaints - for example, Mahadevapura - and constituencies with higher scores had relatively less complaints, on the whole.

The anomalies, such as Govindarajanagar and Vijayanagara, might indicate that the lights in this area have stopped working recently and need to be repaired. These anomalies can be brought to the attention of the city officials through analytics. The constituencies with low scores can be analyzed further with other features to ascertain the causes of the low quality, taking into account the bills of the contractor responsible for installing street lights in the area, the budget allocated to these areas for street lights and the time elapsed since any maintenance on street lights was performed in the ward.

.

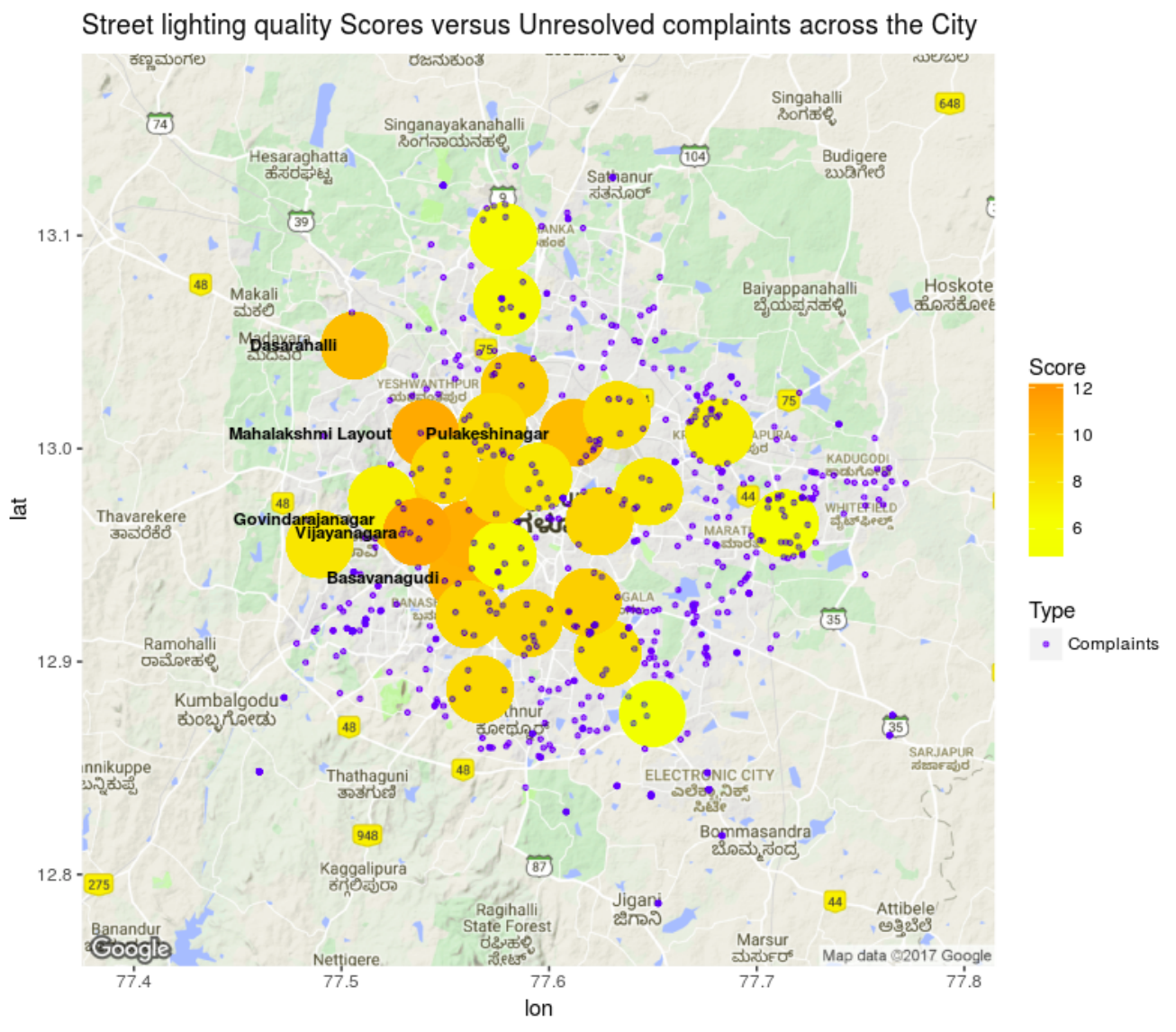

Figure 4. Street lighting quality Vs. Unresolved Complaints across the city.

## IV. Analysis of Citizen Complaints

The complaints data is part of the civic dataset and contains information about the nature and type of all the complaints filed with the BBMP in the fiscal year 2013-14. These categories include a range of categories from Crime and Safety, Maintenance of Roads and Footpaths, Electricity and Power supply, Water services among others. The ward where the complaint was filed is present along with the exact latitude and longitude of the complaint.

Additional attributes of importance are the complaint description, category and state of the work progressed – Resolved, On-the-job or Open. The state of work is further defined as -

- Resolved complaints are one that has been solved by the BBMP.
- On-the-job complaints are ones where workers have been dispatched but the work has not yet been completed.
- Open are complaints which have only been filed, no further action has been done about this complaint.

### A. *Street lighting Quality Complaints*

The first step is to determine how the complaints fall into these three categories to find the efficiency of the BBMP service with respect to citizen complaints. The pie chart in Figure 5 is used to visualize the percentages of complaints in each sector.

The pie chart provides an interesting point of view, contrary to general consensus on the state of work of the BBMP that they never fix things, the number of complaint cases that are not even followed up is only 8%. While the number of complaints that are solved is slightly higher than those still in action with 49% compared to 43%. More clarity would be shed on the resolution efficiency by the addition of a time dimension in the data, measuring the latency in resolution over time.

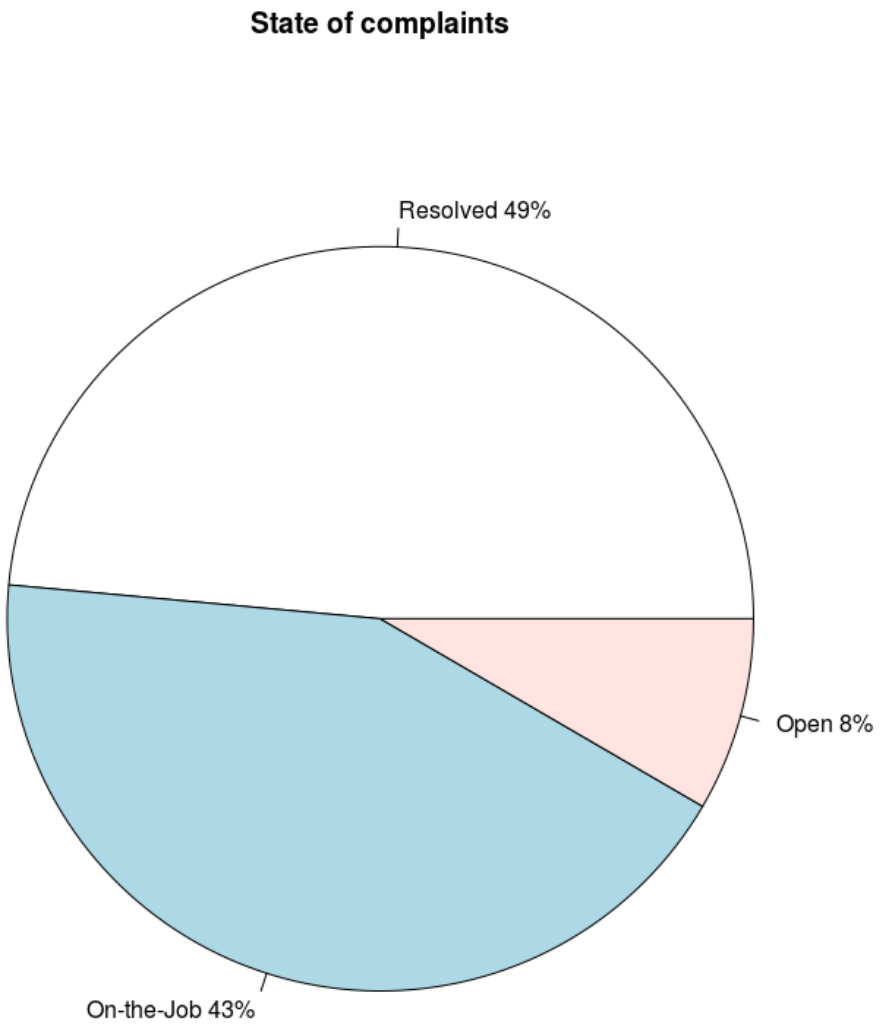

Figure 5. % of Complaints

*B. Complaint Descriptions*

The nature of the complaints vary by ward, with approximately 41,000 rows of complaints. A word cloud for the entire data set could not be generated all at once, due to the tdm matrix requiring upwards of 12GB of memory to create. Hence, complaints were split based on their state of work described by the pie chart. Separate word clouds on each state of the complaint were generated, the result of the resolved state of complaints described in Figure 6.

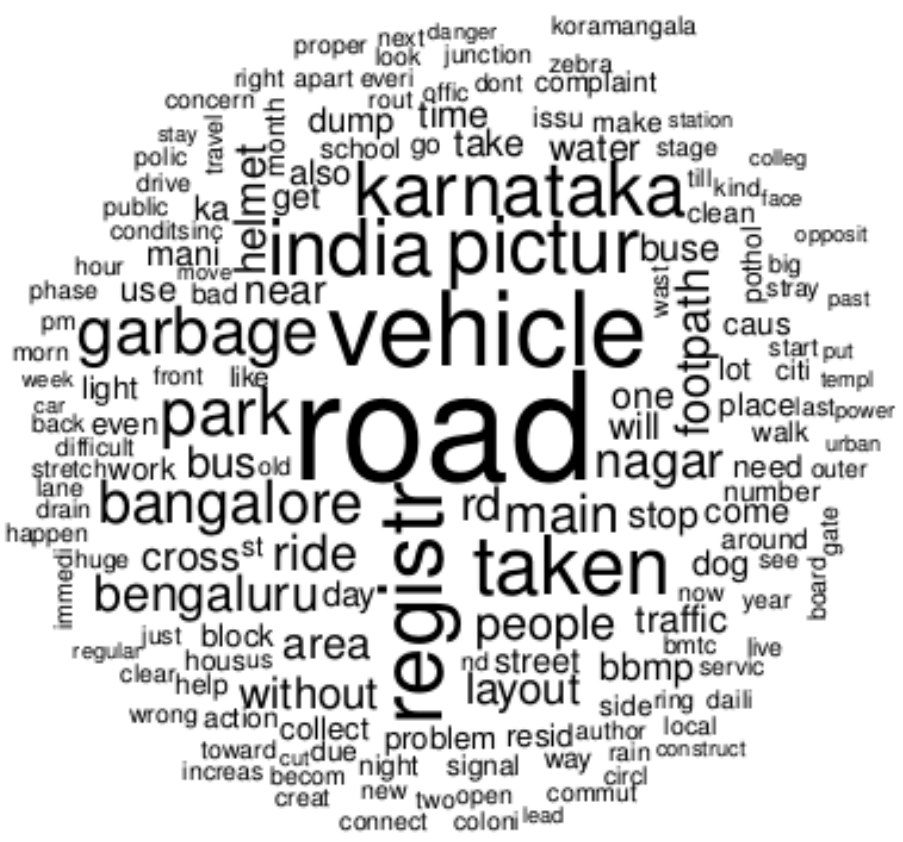

Figure 6. *A Word Cloud of Resolved Complaints*

Having observed the word clouds of all three categories, the most frequent word happens to be "road" in all the three categories of completed, on-the-job and open complaints and in fact dominates by a large margin. This is not surprising given the inadequate connectivity and transport infrastructure, compared to the city population and the rainfall that damages infrastructure seasonally.

To better understand the rest of the complaints, another word cloud was generated, but this time without the word "road". Now there was is a slight difference between the various categories, though major issues such as garbage still come up in all the cases. We now see an increase in footpaths, vehicles, traffic, garbage, bus, water – a diverse set of categories across the city.

## V. Conclusions

Using analytics to assess, improve and monitor progress of cities' is the future of city planning. With the rapid increase in population and urbanization, the amount of civic data generated by a city is too large for humans to sift through to report accurate results. Through data analytics, large amounts of civic data across departments and constituencies can be visualized and analyzed quickly, bringing new insights to the table which may have been overlooked otherwise.

Making the analysis available to the public through various online portals will promote transparency and increase citizen engagement towards building better cities. Through analysis, constituencies which are in need of more amenities can be identified, based on either complaints, population, office/housing spaces etc. Also, citizen satisfaction can be actively measured with respect to the responses of different city departments. Furthermore, city analytics can serve as a performance monitoring system for city corporations in order to increase their efficiency. Harnessing the data for current conditions of the city, complaints and improvements over time with the current budget and contractors can provide insights about the quality of work done by different contractors, the current and forecasted requirements of the city, and can help in allocating the budget in the following years.

The scope of city analytics around the world is expanding rapidly as more data is collected and a growing number of city officials are incorporating analytics into their city improvement and management plans with considerable success.

## VI. Reproducible Research

In the spirit of reproducible research, the work done is publicly available. The R code for the experiments is available here [1].